# Pyrolytic elimination of ethylene from ethoxyquinolines and ethoxyisoquinolines: A computational study

Mohamed A. M. Mahmoud,[1] Mohamed A. Abdel-Rahman,[2] Mohamed F. Shibl,[3*]

[1]Basic Sciences Department, Tanta Higher Institute of Engineering and Technology, Tanta 31511, Egypt.

[2] Chemistry Department, Faculty of Science, Suez University, Suez, 43518, Egypt

[3] Renewable Energy Program, Center for Sustainable Development, College of Arts and Sciences, Qatar University, 2713 Doha

*Corresponding authors:

E-mail: mfshibl@qu.edu.qa (Mohamed F. Shibl)

## ABSTRACT

This work reports thermodynamics and kinetics of unimolecular thermal decomposition of some ethoxyquinolines and ethoxyisoquinolines (1-ethoxyisoquinoline (**1-EisoQ**), 2-ethoxyquinoline (**2-EQ**), 3-ethoxyquinoline (**3-EQ**), 3-ethoxyisoquinoline (**3-EisoQ**), 4-ethoxyquinoline (**4-EQ**), 4-ethoxyisoquinoline (**4-EisoQ**), 5-ethoxyquinoline (**5-EQ**), 5-ethoxyisoquinoline (**5-EisoQ**), 8-ethoxyquinoline (**8-EQ**) and 8-ethoxyisoquinoline (**8-EisoQ**) using density functional theory (BMK, MPW1B95, M06-2X/cc-pvtz) and ab initio (CBS-QB3) calculations. In the course of the decomposition of the investigated systems, ethylene is eliminated with the production of either keto or enol tautomer. The six-membered transition state structure encountered in the path of keto formation is much lower in energy than the four-membered transition state required to give enol form. Rate constants and activation energies for the decomposition of **1-EisoQ**, **2-EQ**, **3-EQ**, **3-EisoQ**, **4-EQ**, **4-EisoQ**, **5-EQ**, **5-EisoQ**, **8-EQ,** and **8-EisoQ** have been estimated at different temperatures and pressures using conventional transition state theory combined with Eckart tunneling (TST/Eck) and the statistical Rice-Ramsperger-Kassel-Marcus (RRKM) theories. The tunneling correction is significant at temperatures up to 1000 K. Rate constants results reveal that ethylene elimination with keto production is favored kinetically and thermodynamically over the whole temperature range of 400-1200 K and the rates of the processes under study increase with the rising of pressure up to 1 bar.

## 1. INTRODUCTION

Quinoline and its derivatives, as important naturally occurring compounds, are found in coal tar as well as bone oil and have biological and pharmaceutical effects [1-9], including antimalarial, antineoplastic, anticonvulsant, antibacterial, antifungal, anticancer, anti-inflammatory, and analgesic activity [1-6]. For the last two decades, nitrogen- and oxygen-containing heterocyclic compounds have been attractive in biology due to their pharmaceutical action, mainly attributable to their ability to make hydrogen bonds. Quinoline research is now one of the most prominent areas in organic, inorganic, pharmaceutical, and theoretical chemistry, as well as dye manufacturing. Tautomerism is a fundamental notion in organic chemistry and an intriguing phenomenon because it is linked to numerous essential chemical and biological processes [10].

The study of how tautomerism affects the chemical, biological, and pharmacological properties of heterocyclic compounds is of great interest to many researchers, particularly medicinal chemists, as it may be related to the pharmacological properties of these compounds. Experimentally and theoretically, the tautomeric equilibrium of heterocyclic compounds has been investigated [11-13], and a detailed analysis of the changes in structural, geometric, and energetic parameters caused by the transfer of hydrogen atoms can help us understand the different properties of tautomers. Understanding the relative stabilities of tautomeric forms of heterocycles and how they convert from one form to another is important in the field of structural chemistry. Quinolines and isoquinolines have been extensively investigated because they are relevant to physics [7], chemistry [8], and medicine [9]. The thermal decomposition of these materials is essential to understand their behavior and stability in different environments [14-23]. Similar to esters [24-27], it was reported that thermolysis of alkoxy benzene and heteroaromatics produces olefins and the corresponding keto or enol form with activation energies depend on the structure of the reactants [14-23]. These gas phase degradation reactions are unimolecular, homogeneous, and pass over six-membered ring transition states [14-23]. In the course of these pyrolytic reactions, different tautomers can be formed. However, the formation of the keto tautomer needs less energy than that required for enol because the former passes over a six-membered ring transition state whereas the latter is formed through a four-membered transition state. Therefore, the keto form appears as a dominant product. If the energy barrier to producing enol from its keto tautomer is low, a state of equilibrium between the two forms might be established.

Experimental [14,16] and theoretical [23] studies have been conducted on the formation of hydroxyquinoline and quinolone by the removal of ethylene from ethoxyquinoline and ethoxyisoquinoline. Al-Awadi and colleagues [14] looked at the rates of thermal ethylene removal from substances such as 2-ethoxyquinoline, 1- and 3-ethoxyisoquinoline, and 1-ethoxythiazole. They also investigated the rates of gas-phase pyrolytic reactions of 2-pyridine and 8-quinoline sulfonic acid esters [16]. Each of the pyridine esters is consistently more reactive than the quinoline ester. This follows from the fact that C-2 of the pyridine will receive a greater electron-withdrawing effect from the nitrogen atom (which will make it the most electron-deficient carbon in the ring), and this indirect electron withdrawal effect of the nitrogen atom in the 2-pyridine esters should facilitate C-O bond cleavage, while for the 8-quinoline esters, this is not the case [16].

Gas phase pyrolysis has been used as a means of organic synthesis [28-35]. For example, flash vacuum pyrolysis of N-alkoxyphthalimides at 673-773 K and 0.02 Torr yielded the corresponding substituted aldehydes and phthalimide [28]. Gas-phase pyrolysis of N-(1H-benzimidazol-2-yl)-N'-arylidenehydrazines produced arylnitriles, 2-aminobenzimid- azole, 2,4,5-triphenylimidazole, 1,3-diphenyl-8H-2,3a,8-triazacyclopenta [a]indene, and 5,11-diphenyl-6H,12H-dibenzimidazo[1,2-a];1',2'-d]pyrazine [29]. Also, the thermal decomposition of 1-(pyrazol-4-yl)-1H-benzotriazole derivatives gave indole and its condensed derivatives [30]. Furthermore, thermolysis of ethyl 2-amino-5-arylazo-6-phenylnicotinates and 5- arylazonicotinonitriles helped in synthesizing some pyridine derivatives [31]. El-Demerdash et al. investigated thermal decompositions of 1-ethoxyisoquinoline (**1-EisoQ**), 2-ethoxyquinoline (**2-EQ**), and 3-ethoxyisoquinoline (**3-EisoQ**) to create ethylene and various tautomers in the gas phase and ethanol solution at the BMK/6-31+G(d,p) and MP2/6-311++G(2d,2p) levels of theory [23]. The thermodynamic and kinetic stability of **2-EQ** and **1-EisoQ** breakdown to ethylene and keto forms is greater than that of the equivalent enols. In the gas phase with ethanol, however, the hydroxy form of **3-EisoQ** is more stable than the keto tautomer [23]. Density functional theory (DFT) and *ab initio* calculations become useful techniques for obtaining information about the structure, relative stability, and other aspects of tautomers, in the sense that quantum chemical calculations can directly examine physical properties of tautomers.

In the present work, thermochemistry and kinetics of the thermal decomposition of 1-ethoxyisoquinoline (**1-EisoQ**), 2-ethoxyquinoline (**2-EQ**), 3-ethoxyquinoline (**3-EQ**), 3-ethoxyisoquinoline (**3-EisoQ**), 4-ethoxyquinoline (**4-EQ**), 4-ethoxyisoquinoline (**4-EisoQ**), 5-ethoxyquinoline (**5-EQ**), 5-ethoxyisoquinoline (**5-EisoQ**), 8-ethoxyquinoline (**8-EQ**) and 8-ethoxyisoquinoline (**8-EisoQ**) to produce ethylene and different tautomers were studied in the gas phase at 400-1200 K and $10^{-6}$ to 10 bar using the BMK/6-31+G(d,p) and MPW1B95/6-311++G(2d,2p), M06-2X/cc-pvtz and ab initio (CBS-QB3) levels of theory. The activation energies and frequency factors of the formed tautomers were described.

## 2. COMPUTATIONAL DETAILS

The density functional theory (DFT) BMK [36] (Boese and Martin) method in conjunction with the 6-31+G(d,p) basis set was employed to optimize reactants, products, and transition states. ChemCraft software V1.8 [37] was used to analyze vibrational frequencies that were scaled with a factor of 0.95 [38]. Based on Hessian matrix analysis, all minima are characterized by having no imaginary frequencies while each transition state comprises only one negative eigenvalue. The located transition states were further verified through minimum energy path (MEP) analysis using intrinsic reaction coordinate (IRC) calculations at the

BMK/ 631+G (d,p) level of theory in mass-weighted Cartesian coordinates [39-41]. The IRC analysis demonstrated that the transition states connect the reactants with their respective products. In addition, to obtain more accurate results, single-point energy calculations were performed using 1-parameter MPW1B95 [42-45], M06-2X [46] functionals and 6-311++G(2d,2p) basis set. MPW1B95 functional has been benchmarked for thermochemistry and kinetics of some unimolecular decomposition reactions [42-45]. MPW1B95 is excellent for broad thermochemistry applications and performs well in hydrogen bonding and weak interaction simulations [42-45]. M06-2X [46,47] functional was used with the correlation consistent cc-pVTZ basis set. Minnesota functional M06-2X [46,47] was developed for accurate thermokinetic investigations. Truhlar and colleagues [46] produced the hybrid meta-generalized gradient M06-2X functional with 54 percent HF exchange-correlation to give accurate kinetic data [46,47]. All main channel frequencies were addressed using the hindered rotor (HR) approximation. One method for locating transition phases is the relaxed scan. Relaxed potential energy scans are performed to obtain the lowest TSs structures achievable.

Energies were refined utilizing the multistep CBS-QB3 [48-50] level at the BMK geometries for accurate chemical kinetic modeling. Low-level calculations on big basis sets, mid-sized sets for second-order correlation corrections, and small basis sets for high-level correlation corrections are all part of the CBS-QB3 composite approach [48-50]. The five-step CBS-QB3 calculation series begins with a geometry optimization at the B3LYP functional using the 6-311++G(2d,2p) basis set, followed by a frequency calculation at the same level to obtain thermal corrections, zero-point vibrational energy, and entropic information. T1 diagnostic calculations of the ROCBS-QB3 [48-50] approach were also employed for TSs and the generated radicals to check the existence or absence of multireference character in the estimated wavefunctions of different species.

The Harmonic Oscillator (HO) approximation provides a direct and fast approach for estimating the vibrational frequencies that calculate the partition function of internal modes, but it fails for large-amplitude internal motions such as internal rotations. By suppressing low-frequency vibrations that correspond to internal rotations, the partition function and thermodynamic characteristics can be calculated using the HO approximation. Because some of the lowest vibrations do not correlate to internal rotations, this poses a severe problem in transition state calculations.

The results were compared with an experiment and accurate cost-effective CBS-QB3 ab initio multilevel computational method [48-50]. Energy barriers calculated for these reactions using this functional and 6-311++G(2d,2p) were recorded as less than 2 kcal/mol. The hindered rotor (HR) technique was utilized to tackle torsional vibration modes in **1-EisoQ**, **2-EQ**, **3-EQ**, **3-EisoQ**, **4-EQ**, **4-EisoQ**, **5-EQ**, **5-EisoQ**, **8-EQ**, **8-EisoQ**, and transition states. However, due to issues with the number of degrees of freedom, we encountered termination faults for several transition states. Rate constants for the successfully calculated TSs with a hampered rotor were determined and compared to those produced using a harmonic oscillator (HO). Very minimal differences were found between them, providing confidence in our conclusions derived utilizing the HO technique. All calculations were conducted with the Gaussian-016W program [51].

The unimolecular thermal decomposition rate constant of **1-EisoQ**, **2-EQ**, **3-EQ**, **3-EisoQ**, **4-EQ**, **4-EisoQ**, **5-EQ**, **5-EisoQ**, **8-EQ,** and **8-EisoQ** were calculated employing the conventional transition state theory (TST) (Eq. (1)) [52-55].

For TST calculations, the dividing surface is located at $s$ = 0.0 and the rate constant reads

$$k^{TST}(T) = \chi(T)\, \sigma \frac{k_B T}{h} \frac{Q^{TS}(T)}{Q^{R}(T)}\, e^{-\frac{V^{\ddagger}(s=0.0)}{k_B T}} \quad (1)$$

where $\chi(T)$ is the tunneling correction, $\sigma$ is the reaction path degeneracy, $k_B$ is the Boltzmann constant, $h$ is the Blank's constant, $T$ is the temperature, $Q^R(T)$ and $Q^{TS}(T)$ are the reactant and transition state partition functions.

One-dimensional (1D) tunneling effects Eckart (Eck) [56] and the fall-off regime by unimolecular Rice-Ramsperger-Kassel-Marcus (RRKM) theory at lower pressures [57-60] are included in TST calculations. Both RRKM and TST theories are implemented in the Kinetic and Statistical Thermodynamical Package (KiSThelP) [61].

Tunneling may play a role in these processes because a hydrogen atom is involved. The tunneling correction is addressed by the Eckart tunneling adjustment in the rate equation. Furthermore, the transmission coefficient $\chi(T)$ was included to account for tunneling along the reaction coordinate, which was estimated using TST and corrected using Eckart tunneling correction factors. Tunneling adjustments are used to correct TST rate coefficients for the asymmetric Eckart's 1D potential energy barrier by integrating the probability, $p(E)$, of transmission across the associated 1D barrier at energy E and the Boltzmann distribution of energies:

$$\kappa_{\text{Eckart}}(T) = \frac{\exp\left(\Delta H_f^{\neq,0K}/k_B T\right)}{k_B T}\int_0^{\infty} p(E)\exp\left(-E/k_B T\right)dE \tag{2}$$

where $\Delta H_f^{\neq,0K}$ represents the zero-point corrected energy barriers in the forward direction.

Using the RRKM method, the microcanonical rate coefficient $k(E)$ is calculated for energy-dependent systems:

$$k(E) = \frac{\sigma G(E)}{hN(E)} \tag{3}$$

where, $\sigma$ is the reaction path degeneracy, $G(E)$ is the total number of states of the transition state with energy less than or equal to $E$, and $N(E)$ is the density of states of the dissociating reactant species. The thermal rate coefficient reads

$$k(T) = \frac{\sigma Q_1^{\ddagger}}{Q_2\, Q_1\, h} e^{-\frac{E_0}{k_b T}} \int_0^{\infty} \frac{G(E) e^{-\frac{E}{k_b T}}}{1+\frac{k\left(E_0+E<\Delta E_j>\right)}{\omega}} dE \tag{4}$$

where, $h$ is Planck's constant, $Q_2$ is the partition function of the active degrees of freedom of the reactant, $Q^{\ddagger}_1$ and $Q_1$ are the partition functions for adiabatic rotations of the transition state and the reactant, respectively, and $E_0$ is the zero-point corrected threshold energy.

Moreover, the strong collision approximation was applied assuming the possible collisions deactivate with $\omega=\beta_c Z_{LJ}[M]$ being the effective collision frequency, where $\beta_c$ represents the collisional efficiency, $Z_{LJ}$ represents the Lennard-Jones collision frequency, and [M] is the total gas concentration. A value of 0.2 was retained for $\beta_c$. Using the Lennard-Jones parameter $\varepsilon/k_B$, where $\varepsilon$ is the energy depth of the Lennard-Jones potential and $\sigma$ represents a dimensionless scale for the molecular radius, we calculated the collision frequencies ($Z_{LJ}$). The Lennard-Jones potential parameters are $\sigma$ = 5.476 Å and $\varepsilon/k_B$ = 367.82 K for EQ whereas $\sigma$ =3.465 Å and $\varepsilon/k_B$ =113.5 K [62] for Argon (Ar) as a diluent gas.

## 3. RESULTS AND DISCUSSION

Figures 1 and 2 show the optimized structures of **1-EisoQ** conformers and their corresponding relative energy, respectively. The SI file depicts all conceivable transition states (TSs) for their unimolecular decomposition in Fig. 3, Fig. 4, and Fig. S1. According to the literature [14,16,66,67] and a relaxed potential scan at MPW1B95, M06-2-X, and the CBS-QB3/BMK/6-31+G(d,p), **1-EisoQ** contains two stable conformers A, B. Inspection of several **1-EisoQ** conformers reveals that the A form is 0.71 kcal mol-1 more stable than the B forms at CBS-QB3/BMK/6-31+G(d,p). As a result, unless otherwise specified, the current study will be based on conformer A.

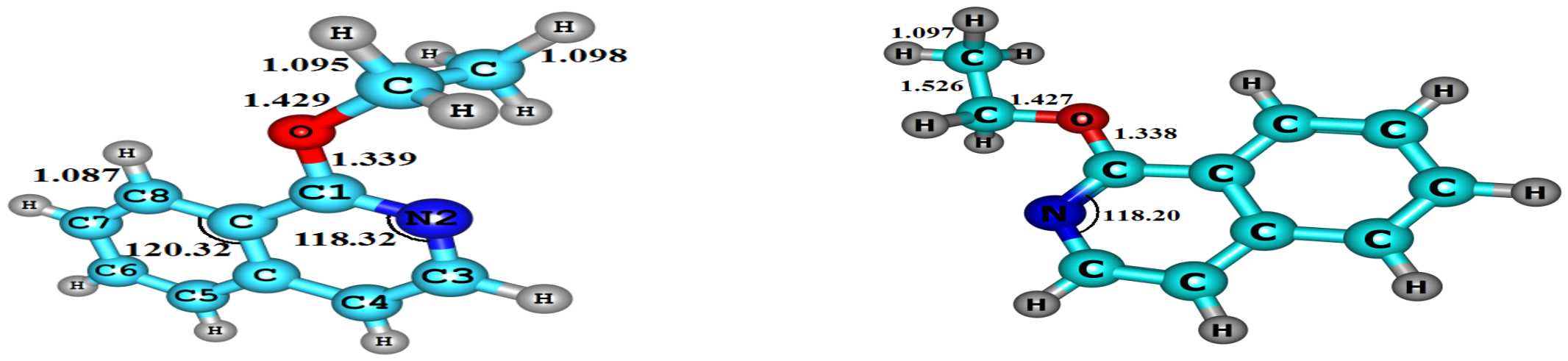

**A** **B**

**Fig. 1.** Optimized structures of **1-EisoQ** conformers (**A**, **B**) at the BMK/6-31+G(d,p) level of theory level.

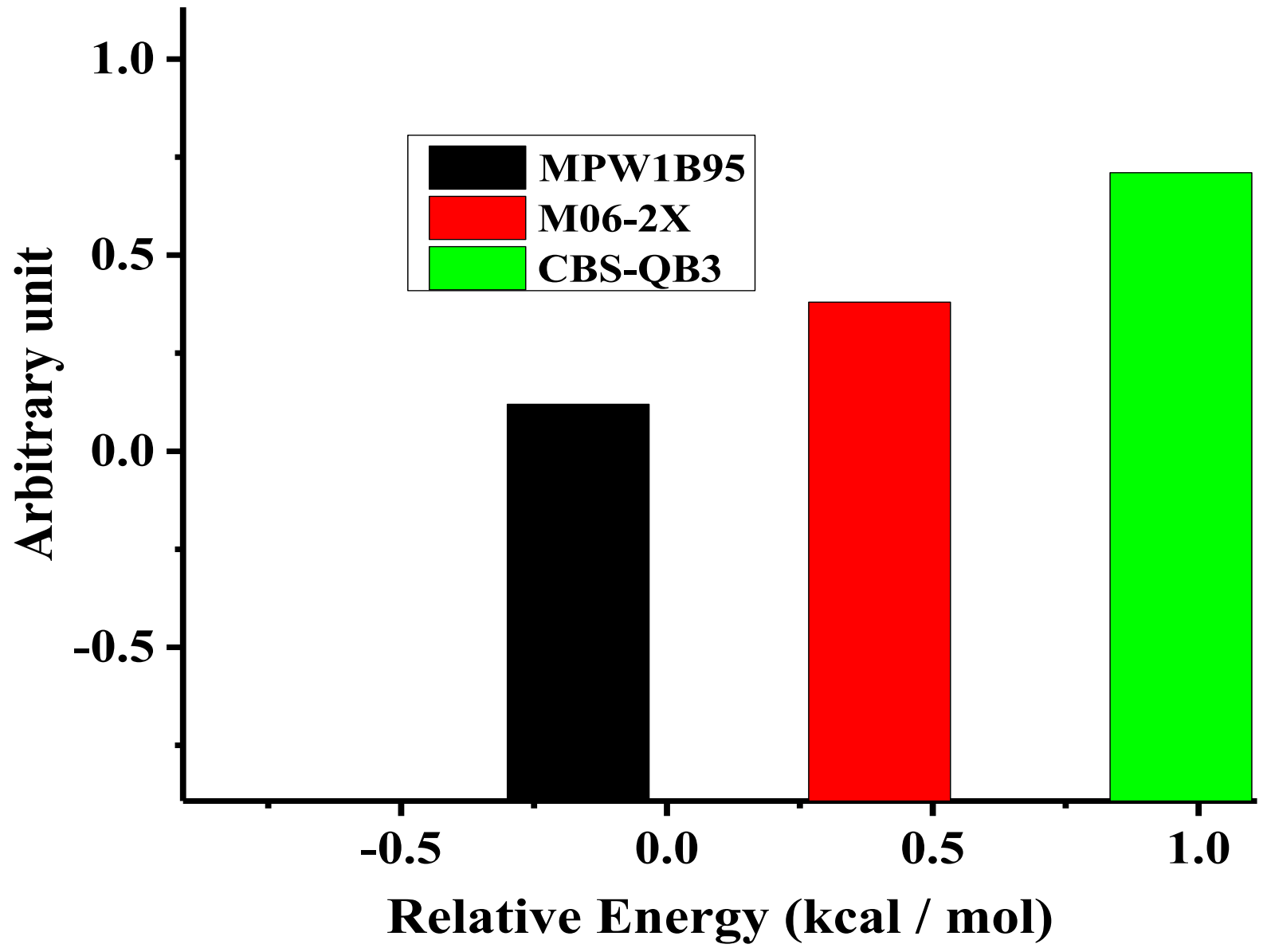

**Fig. 2.** Relative stabilities of **1-EisoQ** conformer (kcal $mol^{-1}$, energies calculated relative to conformer **A**) at MPW1B95, M06-2-X, and the CBS-QB3//BMK/6-31+G(d,p) level.

The unimolecular decomposition reaction of ethoxyquinolines and ethoxyisoquinolines takes place via the breaking of the bond connecting the oxygen atom with the ethyl group along with hydrogen transfer from the $CH_3$ group (1,5- or 1,3-H atom shift) to either nitrogen or oxygen atoms giving rise to keto or enol tautomers with ethylene elimination. The optimized structures of different ethoxyisoquinolines and ethoxyquinolines **(1-EisoQ, 2-EQ, 3-EQ, 3-EisoQ, 4-EQ, 4-EisoQ, 5-EQ, 5-EisoQ, 8-EQ,** and **8-EisoQ)** and their decomposition products are shown in Fig. 3. and Fig.4. The corresponding optimized transition states are displayed in Fig. 3. And Fig.4. The investigated complex bond fission reactions (reactions with barriers) (R1-R13) and simple bond fission reactions (barrierless reactions) (R14-R28) decomposition reactions are summarized as follows:

***(a) Complex bond fission reactions (reactions with barriers)***

**1-EisoQ** → 1-isoQ$_{enol}$ + $C_2H_4$ R1

**1-EisoQ** → 1-isoQ$_{keto}$ + $C_2H_4$ R2

**2-EQ** → 2-Q$_{enol}$ + $C_2H_4$ R3

**2-EQ** → 2-Q$_{keto}$ + $C_2H_4$ R4

**3-EisoQ** → 3-isoQ$_{enol}$ + $C_2H_4$ R5

**3-EisoQ** → 3-isoQ$_{keto}$ + $C_2H_4$ R6

**3-EQ** → 3-$Q_{enol}$ + $C_2H_4$ R7

**4-EisoQ** → 4-iso$Q_{enol}$ + $C_2H_4$ R8

**4-EQ** → 4-$Q_{enol}$ + $C_2H_4$ R9

**5-EisoQ** → 5-iso$Q_{enol}$ + $C_2H_4$ R10

**5-EQ** → 5-$Q_{enol}$ + $C_2H_4$ R11

**8-EisoQ** → 8-iso$Q_{enol}$ + $C_2H_4$ R12

**8-EQ** → 8-$Q_{enol}$ + $C_2H_4$ R13

***Simple bond fission reactions (barrierless reactions)***

R14, 1**-EisoQ** → 1-iso$^{\bullet}$OQ + $^{\bullet}C_2H_5$

R15, 1**-EisoQ** → 1-$^{\bullet}CH_2$OisoQ + $^{\bullet}CH_3$

R16, 1**-EisoQ** → 1-$CH_3{}^{\bullet}$CHOisoQ + $^{\bullet}$H

R17, 1**-EisoQ** → 1-$^{\bullet}CH_2CH_2$OisoQ + $^{\bullet}$H

R18, 1**-EisoQ** → 1-isoQ$^{\bullet}$ + $^{\bullet}OC_2H_5$

R19, **2-EQ** → 2-$^{\bullet}$OQ + $^{\bullet}C_2H_5$

R20, **2-EQ** → 2-$^{\bullet}CH_2$OQ + $^{\bullet}CH_3$

R21, **2-EQ** → 2-$CH_3{}^{\bullet}$CHOQ + $^{\bullet}$H

R22, **2-EQ** → 2-$^{\bullet}CH_2CH_2$OQ + $^{\bullet}$H

R23, **2-EQ** → 2-Q$^{\bullet}$ + $^{\bullet}OC_2H_5$

R24, **3-EisoQ** → 3-$^{\bullet}$OisoQ + $^{\bullet}C_2H_5$

R25, **3-EisoQ** → 3-$^{\bullet}CH_2$OisoQ + $^{\bullet}CH_3$

R26, **3-EisoQ** → 3-$CH_3{}^{\bullet}$CHOisoQ + $^{\bullet}$H

R27, **3-EisoQ** → 3-$^{\bullet}CH_2CH_2$OisoQ + $^{\bullet}$H

R28, **3-EisoQ** → 3-isoQ$^{\bullet}$ + $^{\bullet}OC_2H_5$

The reactions R1-R13 pass through transition states symbolized as $TS1_{enol}$, $TS2_{keto}$, $TS3_{enol}$, $TS4_{keto}$, $TS5_{enol}$, $TS6_{keto}$, $TS7_{enol}$, $TS8_{enol}$, $TS9_{enol}$, $TS10_{enol}$, $TS11_{enol}$, $TS12_{enol}$, and $TS13_{enol}$, respectively.

### 3.1. Enthalpies of Formation

Thermochemical data are critical for comprehending and estimating the stability, reaction pathways, and kinetics of a specific chemical system [61, 63-65]. The computed structures, moments of inertia, vibration frequencies, symmetry, electron degeneracy, and known mass of each molecule are used to calculate entropy and heat capacity contributions as a function of temperature [61, 63-65]. The computational results of fundamental quantities including entropies ($S°_{298}$), enthalpies of formation ($\Delta H_f°_{298}$), and heat capacities ($C_{p298}$) at MPW1B95//BMK/6-31+G(d,p) level are presented in Table 1 for the major species which are depicted in Fig.S.1 in the SI file of **1-EisoQ**, **2-EQ**, **3-EQ,** **3-EisoQ**, **4-EQ,** **4-EisoQ**, **5-EQ**, **5-EisoQ,** **8-EQ**, and **8-EisoQ** pyrolysis. The enthalpies of formation for some selected oxygenates (1-butanol, ethyl propanoate, methyl formate, formic, acetic, and propanoic acids) have been calculated by A.M. El-Nahas et al [45]. A.M. El-Nahas et al [45] found that BMK, MPW195, functionals, and CBS-QB3 have root mean square errors (RMSE) in enthalpies of formation no larger than 4 kcal/mol when compared to the experiment. At 298 K and 1 atm, the $S°_{298}$ and $C_{p298}$ values for each molecule were computed. Even though *Cp* values are independent of the *ab initio* procedure, they are generated from the statistical mechanics treatment [61, 63-65]. Our results demonstrated excellent agreement with earlier experimental [14, 66] and computational [67] results, giving confidence in the unknown thermochemical characteristics of the radicals generated during **1-EisoQ**, **2-EQ**, **3-EQ**, **3-EisoQ**, **4-EQ**, **4-EisoQ**, **5-EQ**, **5-EisoQ**, **8-EQ**, and **8-EisoQ** decomposition.

**Table 1**

Calculated $\Delta H_f°_{298}$ (in kcal mol$^{-1}$), $S°_{298}$ (cal mol$^{-1}$K$^{-1}$), and $C_{p298}$ (cal mol$^{-1}$K$^{-1}$) at MPW1B95//BMK/6-31+G(d,p) level.

| Species | $\Delta H_f°_{298}$ | $S°_{298}$ | $C_{p298}$ | Species | $\Delta H_f°_{298}$ | $S°_{298}$ | $C_{p298}$ |
|---|---|---|---|---|---|---|---|
| **1-EisoQ ($C_{11}H_{11}NO$)** | -6.58 | 24.44 | 10.92 | 3-•$CH_2$OisoQ ($C_{10}H_8NO$) | 41.94 | 23.57 | 9.86 |
| **1-HOisoQ ($C_9H_7NO$)** | -17.33 | 20.93 | 8.48 | 3-$CH_3$•CHOisoQ ($C_{11}H_{10}NO$) | 46.41 | 25.24 | 11.11 |
| **1-isoQO (C9H7NO)** | -24.00 | 24.43 | 10.92 | 3-•$CH_2CH_2$OisoQ ($C_{11}H_{10}NO$) | 38.53 | 25.37 | 11.09 |
| **1-iso•OQ ($C_9H_6NO$)** | 19.57 | 21.38 | 8.26 | 3-isoQ• (C9H6N) | 90.41 | 20.07 | 7.41 |
| **1-•CH2OisoQ ($C_{10}H_8NO$)** | 38.94 | 23.22 | 9.93 | 3-EQ ($C_{11}H_{11}NO$) | 8.22 | 24.56 | 10.98 |
| **1-$CH_3$•CHOisoQ ($C_{11}H_{10}NO$)** | 43.35 | 24.94 | 11.04 | 3-HOQ ($C_9H_7NO$) | -7.16 | 21.17 | 8.67 |
| **1-•$CH_2CH_2$OisoQ ($C_{11}H_{10}NO$)** | 35.69 | 25.16 | 11.10 | 4-EisoQ ($C_{11}H_{11}NO$) | 6.25 | 24.41 | 10.94 |
| **1-isoQ• ($C_9H_6N$)** | 87.50 | 20.09 | 7.42 | 4-isoHOQ ($C_9H_7NO$) | -4.53 | 22.99 | 8.67 |
| **2-EQ ($C_{11}H_{11}NO$)** | -65.57 | 24.22 | 10.91 | 4-EQ ($C_{11}H_{11}NO$) | 3.49 | 24.73 | 10.97 |
| **2-HOQ ($C_9H_7NO$)** | -18.41 (34.63, 45.20 Exp) [14] (4.3,-17.8,-28.7) [67] | 20.96 | 8.54 | 4-HOQ ($C_9H_7NO$) | -8.98 (27.32, 38.79 Exp) [14] (16.23, 19.53Theo [67], 20,8 Exp) [66] | 21.43 | 8.65 |
| | (-17.80,4,60 Theo,-25.5 Exp) [66] | | | | | | |
| **2-QO ($C_9H_7NO$)** | -23.10 | 21.06 | 8.51 | 5-EisoQ ($C_{11}H_{11}NO$) | 5.75 | 24.46 | 10.94 |
| **2-•OQ ($C_9H_6NO$)** | 18.53 | 21.36 | 8.32 | 5-isoHOQ ($C_9H_7NO$) | -5.96 | 22.28 | 8.75 |
| **2-$CH_3$•CHOQ ($C_{10}H_8NO$)** | 37.84 | 23.14 | 9.93 | 5-EQ ($C_{11}H_{11}NO$) | 4.53 | 24.42 | 10.95 |
| **2-$CH_3$•CHOQ ($C_{11}H_{10}NO$)** | 45.62 | 24.86 | 11.09 | 5-HOQ ($C_9H_7NO$) | -6.25 | 21.490 | 8.69 |
| **2-•$CH_2CH_2$OQ ($C_{11}H_{10}NO$)** | 39.07 | 25.02 | 11.06 | 8-EisoQ ($C_{11}H_{11}NO$) | -1.40 | 24.58 | 10.98 |
| **2-Q• ($C_9H_6N$)** | 83.21 | 20.04 | 7.42 | 8-isoHOQ ($C_9H_7NO$) | -5.67 | 21.50 | 8.67 |
| **3-EisoQ ($C_{11}H_{11}NO$)** | 1.27 | 24.40 | 10.91 | 8-EQ ($C_{11}H_{11}NO$) | 5.39 | 24.41 | 10.97 |
| **3-HOisoQ ($C_9H_7NO$)** | -13.84 | 21.02 | 8.55 | 8-HOQ ($C_9H_7NO$) | -13.41 (19.42, 19.84Exp) [14] (1.94, 34.36 Theo[67], 6.5 Exp) [66] | 20.87 | 8.48 |
| **3-isoQO ($C_9H_7NO$)** | -9.59 | 21.09 | 8.54 | 8-QO ($C_9H_7NO$) | 2.34 | 20.94 | 8.43 |
| **3-•OisoQ ($C_9H_6NO$)** | 21.51 | 21.40 | 8.27 | | | | |

Ref, Exp. Al-Awadi NA et al [14], Ref, Exp. M. A.V. Silva et al [66] and Ref, Theo. M.M. Meshhal et al [67].

## 3. 1. ENERGETICS

Enthalpy profiles for the unimolecular decomposition of **1-EisoQ**, **2-EQ**, **3-EQ, 3-EisoQ**, **4-EQ, 4-EisoQ**, **5-EQ**, **5-EisoQ, 8-EQ**, and **8-EisoQ** at MPW1B95/6-311++G (2d,2p)// BMK/6-31+G(d,p) are illustrated in Fig. 5. and Fig. 6. Gibbs free energies for the investigated reactions calculated at the same level of theory are collected in Table 2. Keto formation passes through six-membered ring transition states while producing the corresponding enol tautomer is accomplished via a four-membered ring transition state. Therefore, the former reaction requires less energy than the latter as a result of the stability of the six-membered ring relative to the four-membered transition state. For example, the production of keto and enol tautomers from **1-isoQ$_{enol}$** needs free energy barriers of 35.0 and 51.0 kcal/mol, respectively with reaction enthalpy change of 4.9 and 10.5 kcal/mol. Therefore, the decomposition of **1-EisoQ** to yield the keto form is thermodynamically and kinetically more preferable than the formation of the enol tautomer. Similarly, the formation of 2-quinolone (**2-Q$_{keto}$**) is kinetically and thermodynamically more favorable compared to its enol (**2-Q$_{enol}$**), 35.6, 50.6, 6.5, and 10.6 kcal/mol, respectively. Decomposition of **8-EQ** to enol form and ethylene is the least endothermic channel and most thermodynamically favored reaction with a higher degree of spontaneity (free energy change equals -9.86 kcal/mol) where the hydrogen bond between the hydrogen atom of the hydroxyl group with the nitrogen atom plays a significant role. This is clear when we compare the endothermicity and spontaneity of forming 8-hydroyisoquinoline with that of 8-hydroxyquinoline, see Table 2.

When enols are formed from ethoxyquinoline or ethoxyisoquinoline where the ethoxy group is not adjacent to the nitrogen atom (**3-EQ**, **4-EQ**, **4-EisoQ**, **5-EQ**, **5-EisoQ**, **8-EQ**, and **8-EisoQ**), the energy barrier for 1,3-H atom shift is lowered by ca. 3-5 kcal/mol. The same finding has been noticed in the H-atom shift in a series of some six-membered carbo- and heterocyclic compounds [68].

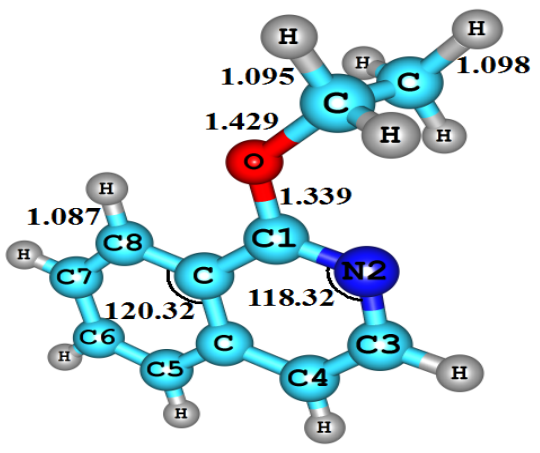

**1-EisoQ**

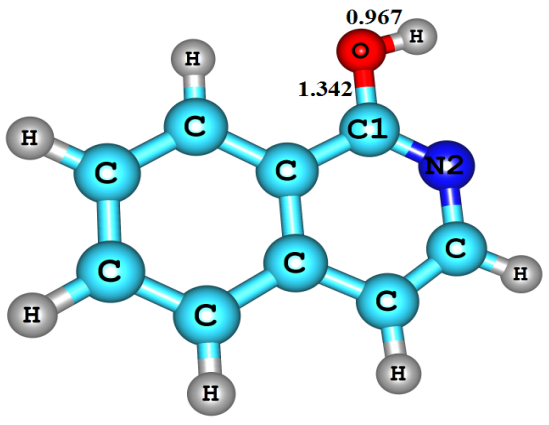

**1-isoQ$_{enol}$ (P1)**

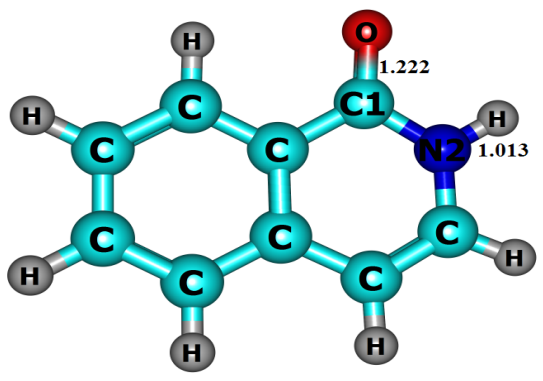

**1-isoQ$_{keto}$ (P2)**

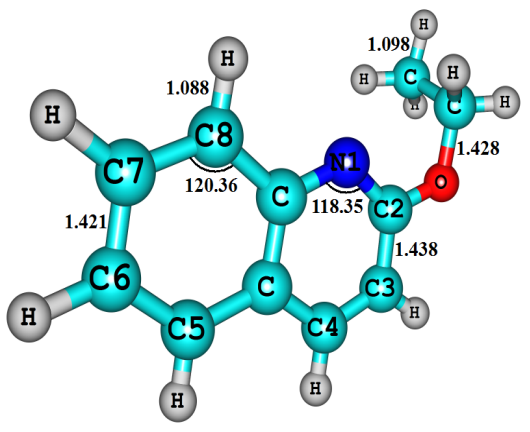

**2-EQ**

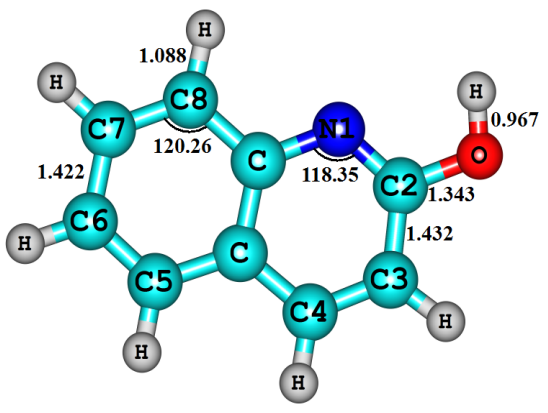

**2-Q$_{enol}$ (P3)**

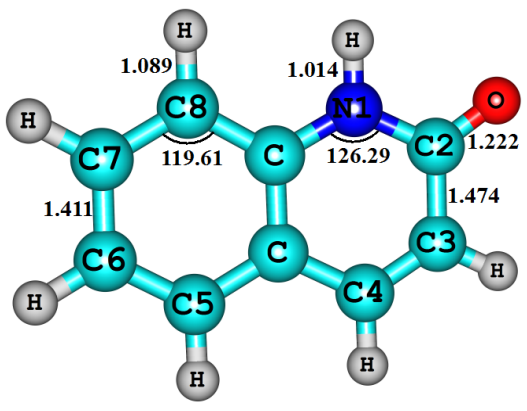

**2-Q$_{keto}$ (P4)**

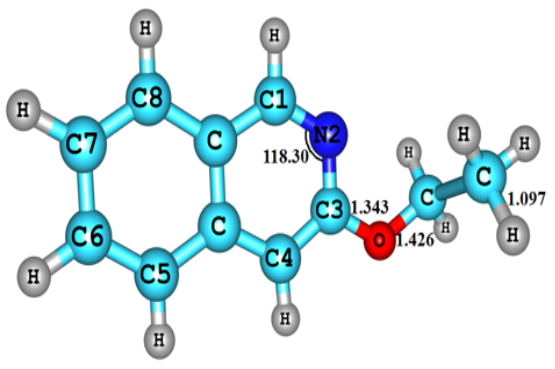

**3-EisoQ**

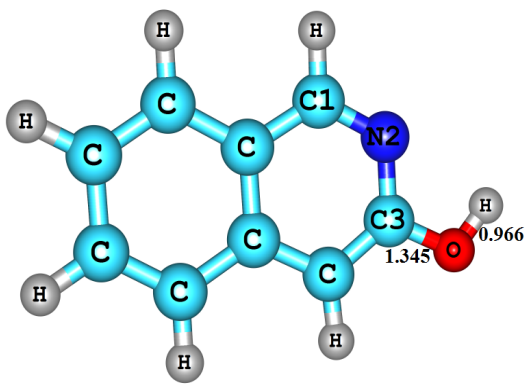

**3-isoQ$_{enol}$ (P5)**

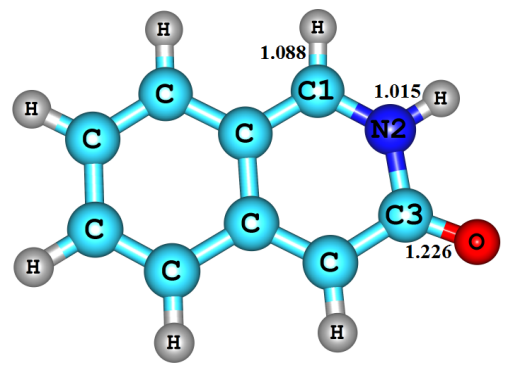

**3-isoQ$_{keto}$ (P6)**

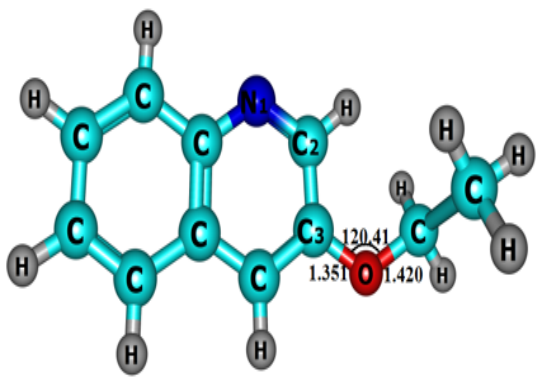

**3-EQ**

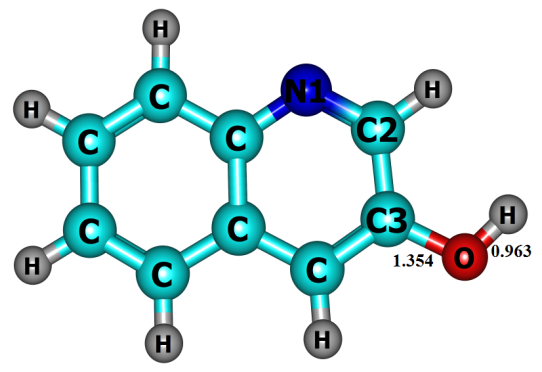

**3-Q$_{enol}$ (P7)**

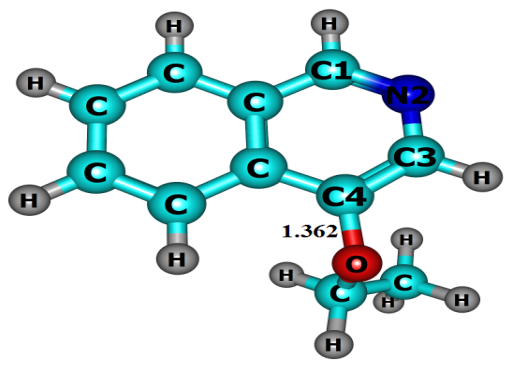

**4-EisoQ**

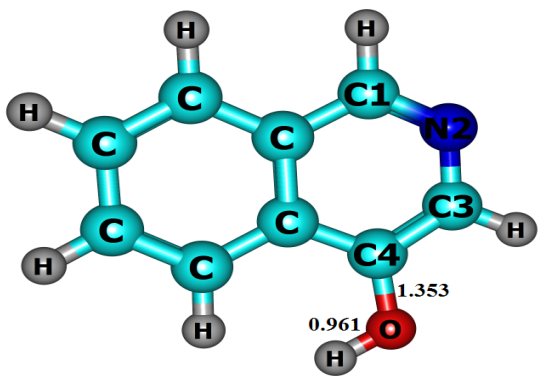

**4-isoQ$_{enol}$ (P8)**

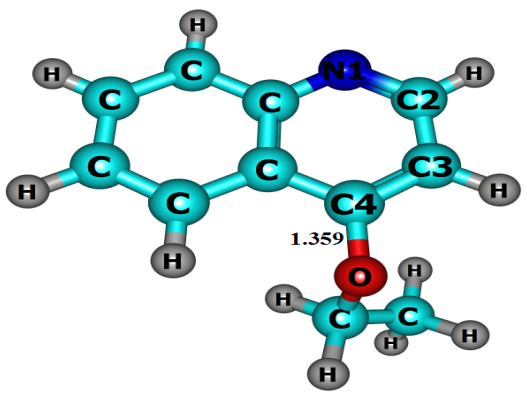

**4-EQ**

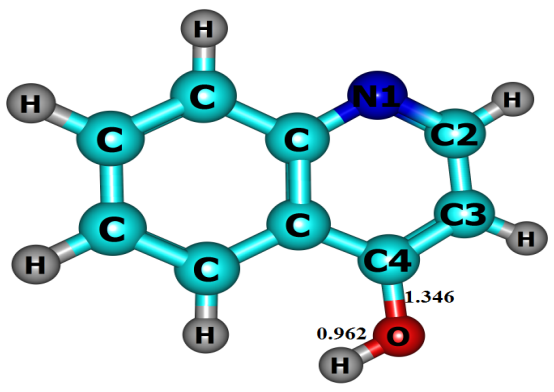

**4-Q$_{enol}$ (P9)**

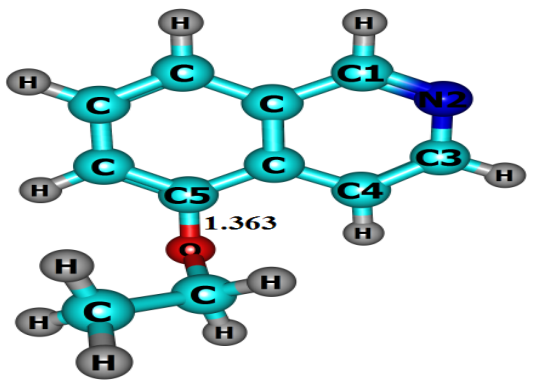

**5-EisoQ**

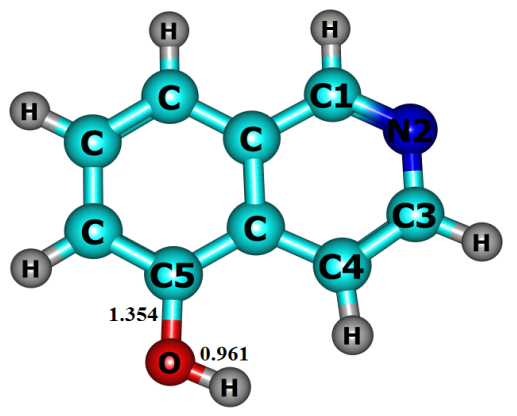

**5-isoQ$_{enol}$ (P10)**

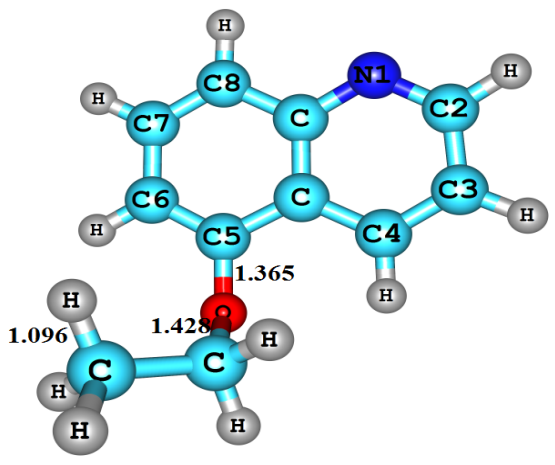

**5-EQ**

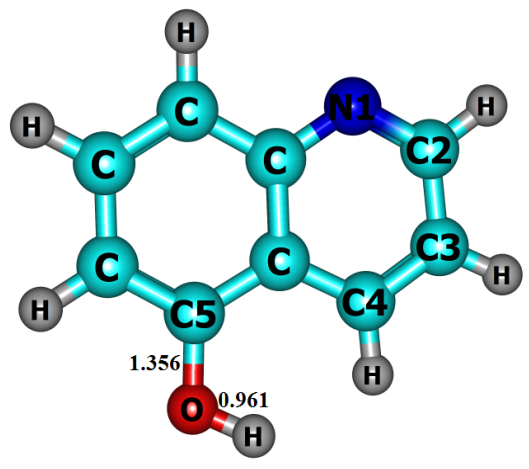

**5-Q$_{enol}$ (P11)**

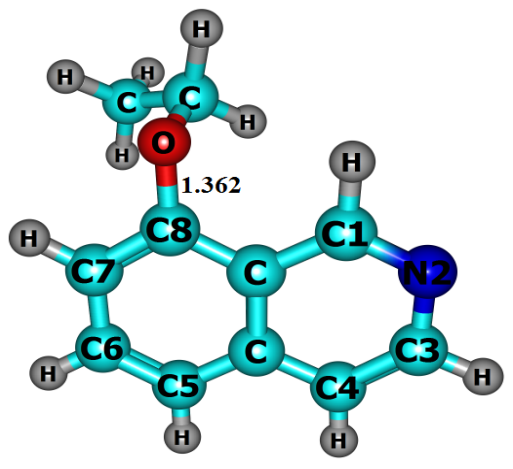

**8-EisoQ**

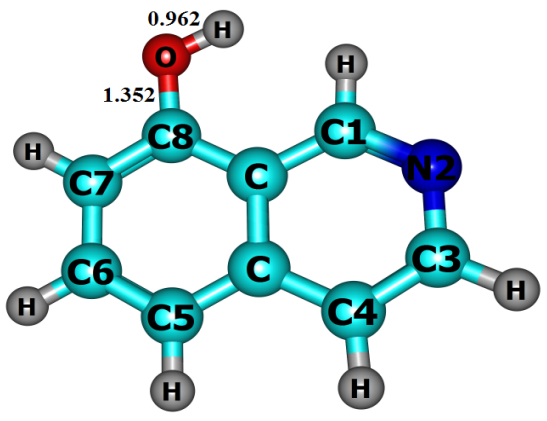

**8-isoQ$_{enol}$ (P12)**

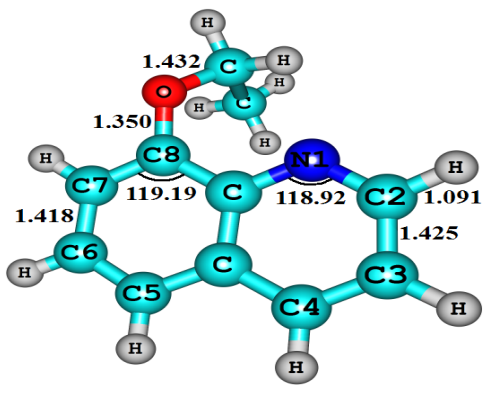

**8-EQ**

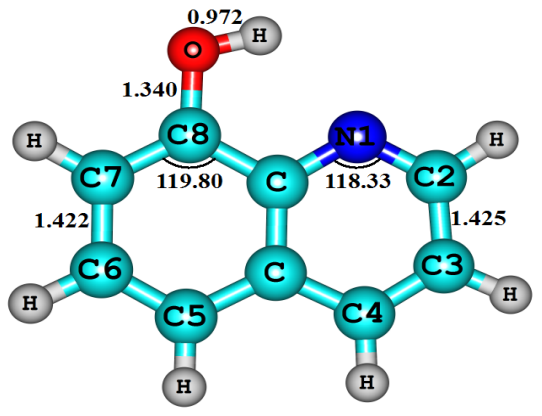

**8-Q$_{enol}$ (P13)**

**Fig. 3.** Optimized structures of ethoxyisoquinolines, ethoxyquinolines, and their corresponding decomposition products calculated at BMK/6-31+G(d,p) level of theory. Bond lengths and angles are given in Ångström and degree, respectively.

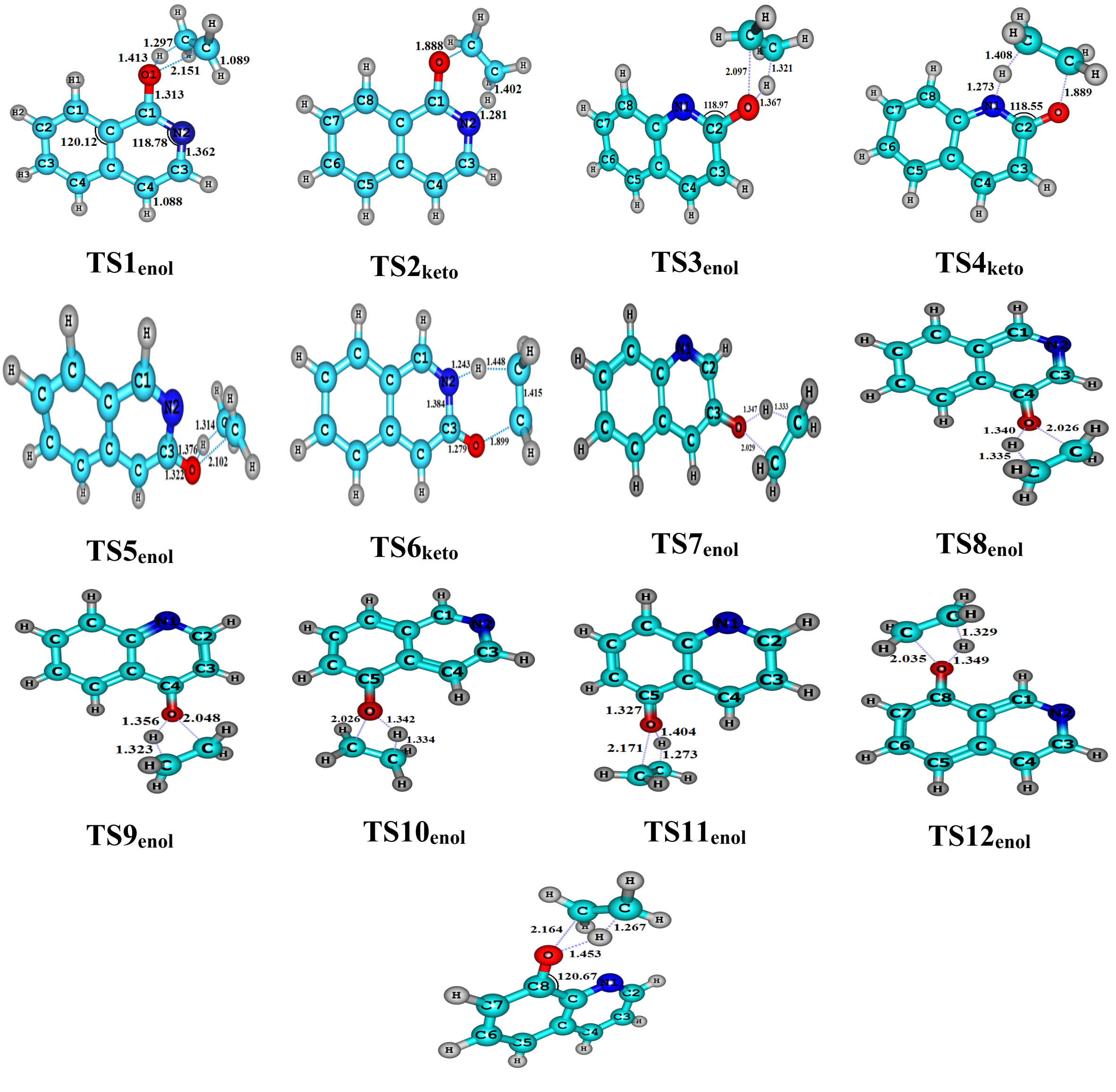

**Fig. 4.** Optimized transition state structures for the thermal decomposition reactions of ethoxyisoquinolines, ethoxyquinolines calculated at BMK/6-31+G(d,p). Bond lengths and angles are given in Ångström and degree, respectively.

(a) Complex bond fission reactions

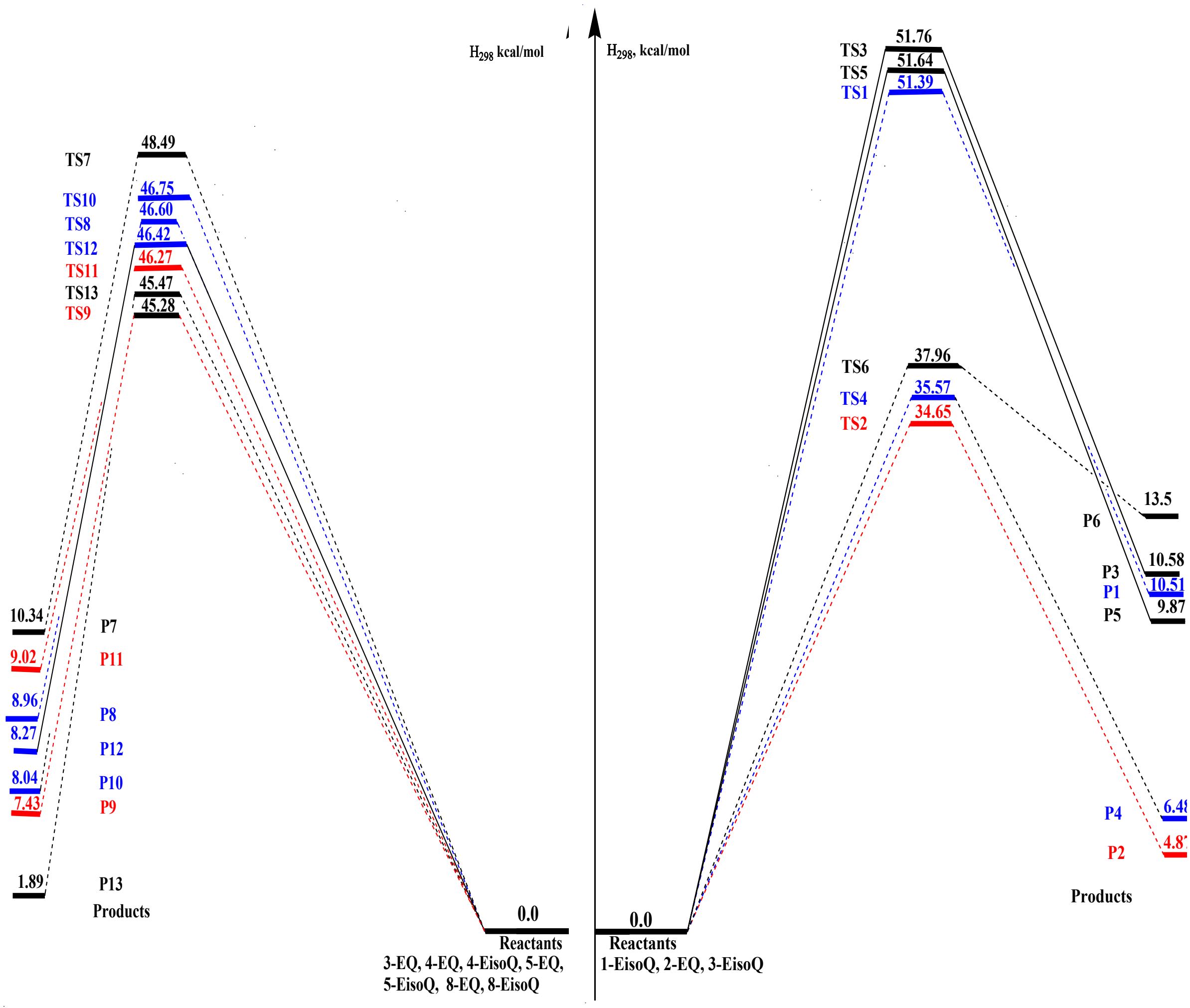

**Fig. 5.** The enthalpy profile (in kcal/mol) for unimolecular decomposition reactions of the investigated ethoxyquinolines and ethoxyisoquinolines (R1- R13, Pn=keto or enol + ethylene) at MPW1B95/6-311++G(2d,2p)//BMK/6-31+G(d,p).

(b) Simple bond fission reactions

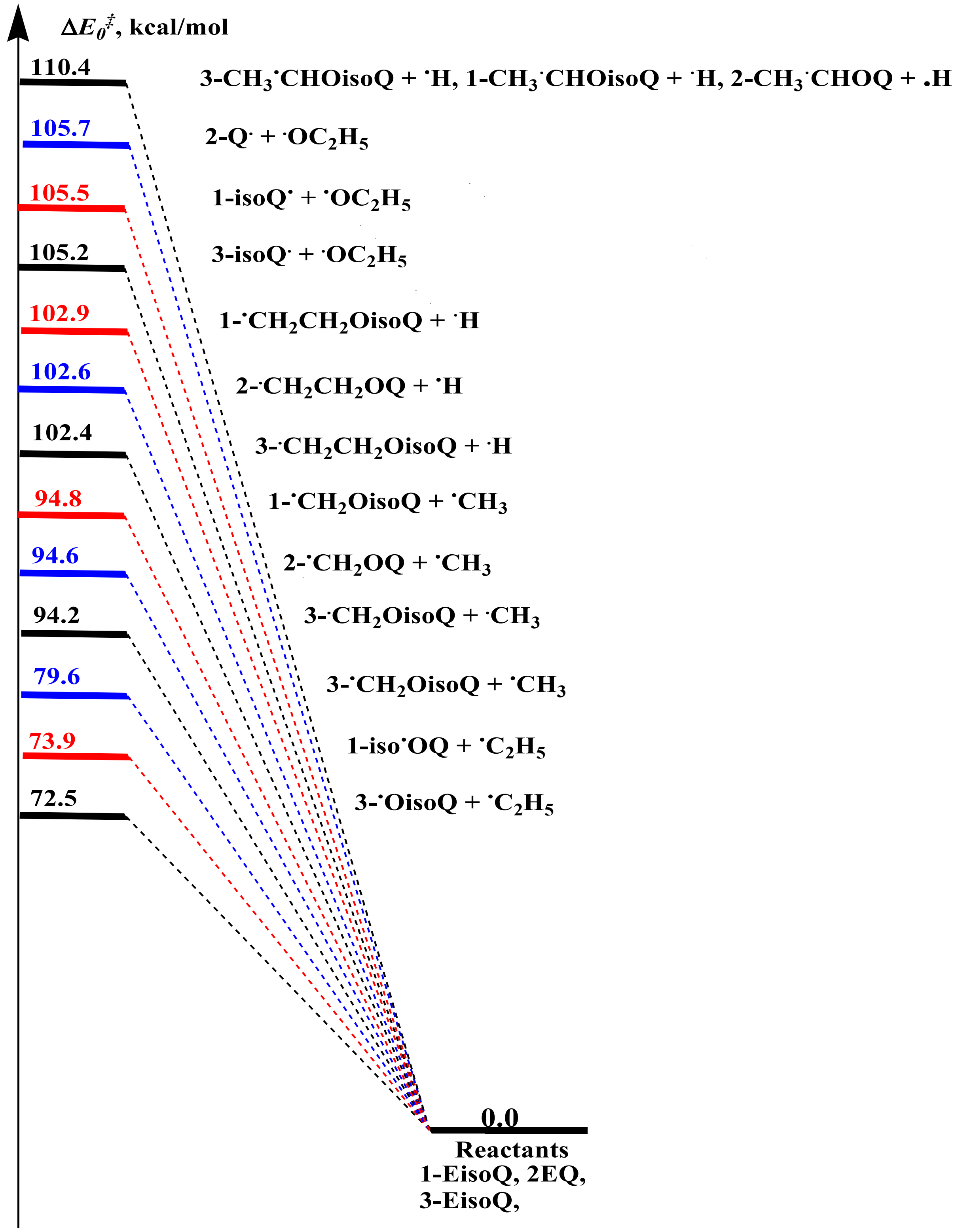

**Fig. 6.** Potential energy profile for unimolecular decomposition of dominant reactions **1-EisoQ**, **2-EQ**, **3-EisoQ**, **(**$\Delta E_0$, $\Delta E_0^{‡}$, kcal/mol) using MPW1B95/6-311++G(2d,2p)//BMK/6-31+G(d,p) level of theory.

## 3. 2. RATE CONSTANTS

Rate constants ($k_{1-13}$) for all reactions are shown in Figs. 7-11 and Table 3. They were calculated using the conventional transition state theory (TST) combined with Eckart tunneling (TST/Eck) as well as the statistical Rice-Ramsperger-Kassel-Marcus (RRKM) theories over a temperature range of 400-1200 K at 1 bar. The employed temperature range was selected for two reasons: (i) it covers the experimental kinetic measurements; (ii) the rates at this temperature range allow comparisons among the present systems and other related compounds.

**Table 2**

Zero-point corrected relative energies, enthalpies, and free energies (in kcal/mol) for unimolecular decomposition for the investigated systems at different DFT (MPW1B95, M06-2X/cc-pvtz): DFT//BMK/6-31+G(d,p) and CBS-QB3 methods (P = 1 bar, T =298 K).

| Parameters Species | $\Delta E_0^{‡}$ | $\Delta H^{‡}_{298}$ | $\Delta G^{‡}_{298}$ | Products | $\Delta E_0$ | $\Delta H_{298}$ | $\Delta G_{298}$ |
|---|---|---|---|---|---|---|---|
| **1-EisoQ** | 0.00 | 0.00 | 0.00 | | 0.00 | 0.00 | 0.00 |
| **TS1$_{enol-cis}$** | 51.24, **65.45,** *64.18* | 51.39, **65.60,** *64.40* | 51.00, **65.21,** *64.18* | **P1, 1-isoQ$_{enol}$** | 9.78, **15.72,** *14.63* | 10.51, **16.45,** *15.45* | -1.28, **4.65,** *3.39* |
| **TS1$_{enol-trans}$** | 51.13, **65.07,** *64.89* | 51.19, **65.14,** *64.84* | 51.57, **64.99,** *65.12* | | 10.19, **15.34,** *6.62* | 10.31, **16.42***, 15.89* | -0.71, **4.43,** *4.69* |
| **TS2$_{keto-cis}$** | 34.84, **47.54,** *46.06* | 34.65, **47.34,** *45.94* | 35.12, **47.82,** *46.21* | **P2, 1-isoQ$_{keto}$** | 4.10, **11.53,** *10.27* | 4.87, **12.29,** *11.09* | -7.04, **0.39***, -1.03* |
| **TS2$_{keto-trans}$** | 34.72, **47.16,** *46.77* | 34.45, **46.88,** *46.38* | 35.7, **47.6,** *47.52* | | 4.52, **11.15,** *2.26* | 4.67, **12.27,** *11.54* | -6.47, **0.17,** *0.27* |
| **2-EQ** | 0.00 | 0.00 | 0.00 | | 0.00 | 0.00 | 0.00 |
| **TS3$_{enol}$** | 51.38, **65.23,** *64.68* | 51.76, **65.61,** *64.88* | 50.57, **64.42,** *64.27* | **P3, 2-Q$_{enol}$** | 9.73, **15.56,** *14.79* | 10.58, **16.42,** *15.61* | -1.5, **4.33,** *3.51* |
| **TS4$_{keto}$** | 35.59, **48.12,** *47.04* | 35.57, **48.10,** *46.93* | 35.58, **48.11,** *47.23* | **P4, 2-Q$_{keto}$** | 5.58, **13.09,** *18.32* | 6.48, **13.98,** *19.19* | -5.76, **1.74,** *0.66* |
| **3-EisoQ** | 0.00 | 0.00 | 0.00 | | 0.00 | 0.00 | 0.00 |
| **TS5$_{enol}$** | 51.41, **65.66,** *64.68* | 51.64, **65.90,** *64.9* | 51.04, **65.29,** *64.32* | **P5, 3-isoQ$_{enol}$** | 9.04, **14.76,** *13.67* | 9.87, **15.59,** *14.4* | -2.07, **3.66,** *2.5* |
| **TS6$_{keto}$** | 37.99, **51.98,** *50.94* | 37.96, **51.95,** *50.83* | 38.10, **52.09,** *51.20* | **P6, 3-isoQ$_{keto}$** | 12.65, **22.71,** *21.1* | 13.50, **23.57,** *21.91* | 1.48, **11.54,** *9.89* |
| **3-EQ** | 0.00 | 0.00 | 0.00 | | 0.00 | 0.00 | 0.00 |
| **TS7$_{enol}$** | 48.32, **77.93,** *75.96* | 48.49, **78.10,** 75.67 | 47.88, **77.49,** *76.6* | **P7, 3-Q$_{enol}$** | 9.45, **15.07,** *13.92* | 10.34, **15.97,** *14.91* | -1.59, **4.04,** *2.71* |
| **4-EisoQ** | 0.00 | 0.00 | 0.00 | | 0.00 | 0.00 | 0.00 |
| **TS8$_{enol}$** | 46.50, **61.94,** *60.57* | 46.60, **62.04,** *60.72* | 46.19, **61.63,** *60.21* | **P8,4-isoQ$_{enol}$** | 7.81, **15.40,** *14.21* | 8.96, **16.55,** *14.86* | -5.42, **2.17,** *3.17* |
| **4-EQ** | 0.00 | 0.00 | 0.00 | | 0.00 | 0.00 | 0.00 |
| **TS9$_{enol}$** | 45.19, **60.44,** *58.95* | 45.28, **60.53,** *59.1* | 45.04, **60.30,** *58.65* | **P9,4-Q$_{enol}$** | 6.52, **13.87,** *12.85* | 7.43, **14.78,** *13.82* | -4.62, **2.73,** *1.53* |
| **5-EisoQ** | 0.00 | 0.00 | 0.00 | | 0.00 | 0.00 | 0.00 |
| **TS10$_{enol}$** | 46.64, **61.83,** *60.54* | 46.75, **61.95,** *60.7* | 46.29, **61.48,** *60.1* | **P10,5-isoQ$_{enol}$** | 6.84, **14.37,** *13.79* | 8.04, **15.56,** *14.88* | -5.39, **2.13,** *2.22* |
| **5-EQ** | 0.00 | 0.00 | 0.00 | | 0.00 | 0.00 | 0.00 |
| **TS11$_{enol}$** | 45.98, **64.33,** *61.37* | 46.27, **64.62,** *61.67* | 44.92, **63.27,** *60.72* | **P11, 5-Q$_{enol}$** | 8.01, **15.54,** *14.39* | 9.02, **16.55,** *15.***53** | -3.36, **4.16,** *2.64* |
| **8-EisoQ** | 0.00 | 0.00 | 0.00 | | 0.00 | 0.00 | 0.00 |
| **TS12$_{enol}$** | 46.28, **61.48,** *60.21* | 46.42, **61.61,** *60.36* | 45.94, **61.13,** *59.85* | **P12,8-isoQ$_{enol}$** | 7.30, **14.86,** *13.86* | 8.27, **15.82,** *14.89* | -4.04, **3.52,** *2.41* |
| **8-EQ** | 0.00 | 0.00 | 0.00 | | 0.00 | 0.00 | 0.00 |
| **TS13$_{enol}$** | 45.72, **60.62**, *59.28* | 45.47, **60.37**, *59.48* | 44.86, **59.75**, *59.15* | **P13, 8-Q$_{enol}$** | 1.18, **7.92**, *7.15* | 1.89, **8.74**, *15.28* | -9.86, **-3.01**, *-3.89* |

Detailed rate constants calculated from TST and RRKM theories along with tunneling corrections (Eck) for the H-atom transfer reactions (R1–R13) at the MPW1B95//BMK/6-31+G(d,p) at 400-1200 K and 1 bar are listed in the supporting information (SI, Tables S1-S13). At a low temperature of 400 K, Eck tunneling gives higher contributions of 5.69, 5.97, 6.02, 7.15, 6.50, 5.32, 6.60, and 6.06 for R1, R3, R5, R7, R8, R9, R10, and R12, respectively, than other reactions. In general, the contribution of tunneling is marginal for all reactions at temperatures ≥ 1000 K. The tunneling contribution decreases with the rising of temperature until reaching 1.2 at T > 1000 K. Plots of TST/Eck rate constants ($k_1$-$k_{13}$) against temperature as depicted in Figs. 7 -10 illustrate an Arrhenius behavior. Inspection of these Figures reveals positive temperature dependence with rate coefficients (frequency factor ($A$) and activation energy ($E_a$)) fit with two-parameter equations ($k = A\ exp^{(-Ea/RT)}$).

Figs. 8 and 9 show a comparison of Arrhenius plots for the main dominant pathways of reaction decomposition (R1, R3, R5) in this study at 600 K and 1 bar and from Al-Awadi et al [14,16]. The results indicate comparable results of rate constants.

The calculated $A$ and $E_a$ values as derived from the displayed Arrhenius plots and collected in Table 3 are in reasonable agreement with the reported experimental data of Al-Awadi et al [14-22] for quinolines, isoquinolines, and other aromatic systems bearing ethoxy group.

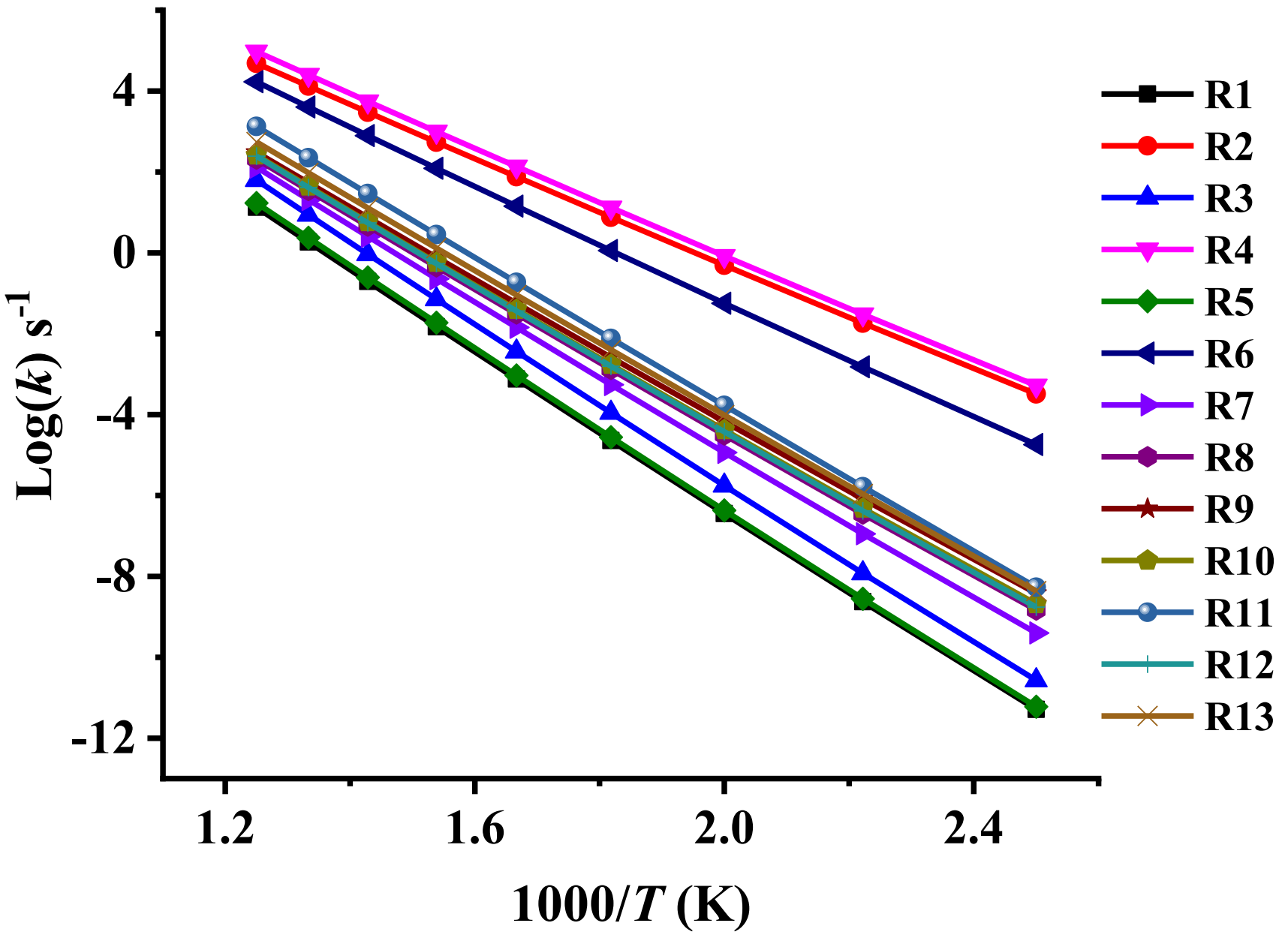

**Fig. 7.** TST/Eck rate constants for unimolecular decomposition reactions of the investigated ethoxyquinolines and ethoxyisoquinolines (R1-R13) at ($T$ = 400-1200 K, $p$ =1 bar) calculated at MPW1B95/6-311++G(2d,2p)//BMK/6-31+G(d,p) level of theory.

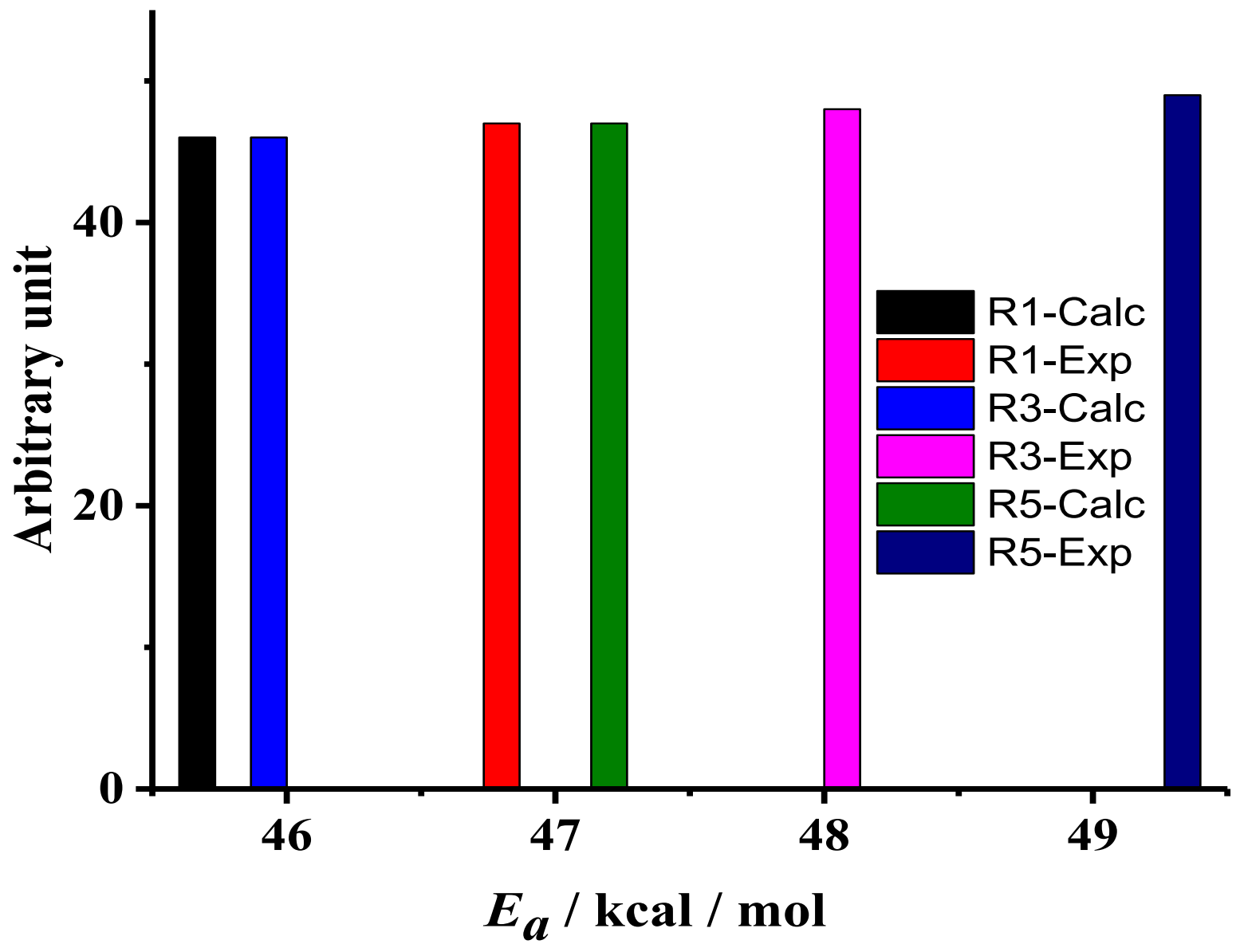

**Fig. 8.** Comparison of the calculated high-pressure limit rate coefficients ($E_a$/kcal/mol) dominant reactions (R1, R3, R5) at MPW1B95/6-311++G(2d,2p)//BMK/6-31+G(d,p) level of theory with Exp., Al-Awadi et al [14,16].

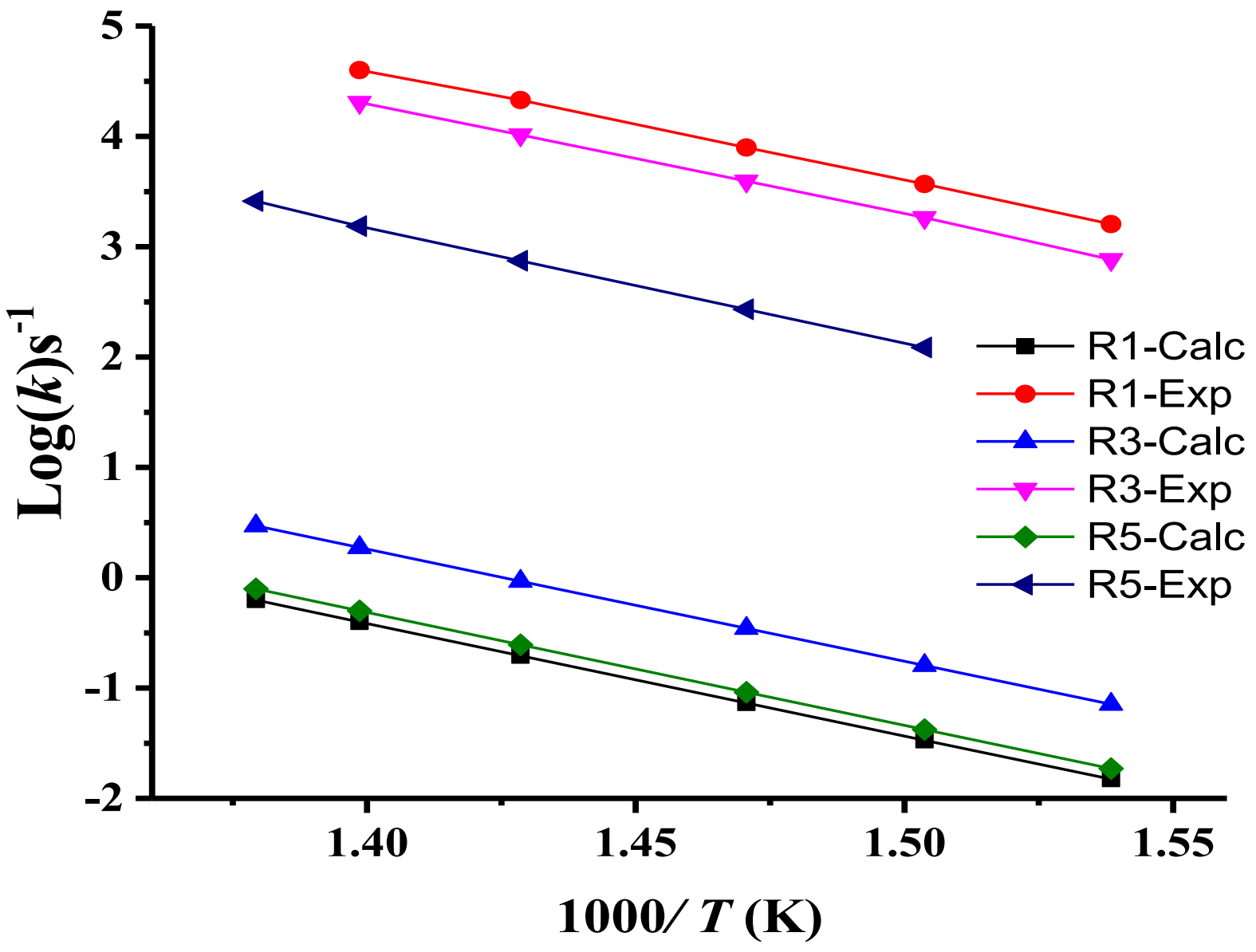

**Fig. 9.** Comparison of the calculated high-pressure limit rate coefficients (Log (*k*)s$^{-1}$) dominant reactions (R1, R3, R5) at MPW1B95/6-311++G(2d,2p)//BMK/6-31+G(d,p) level of theory with Exp., Al-Awadi et al [14,16].

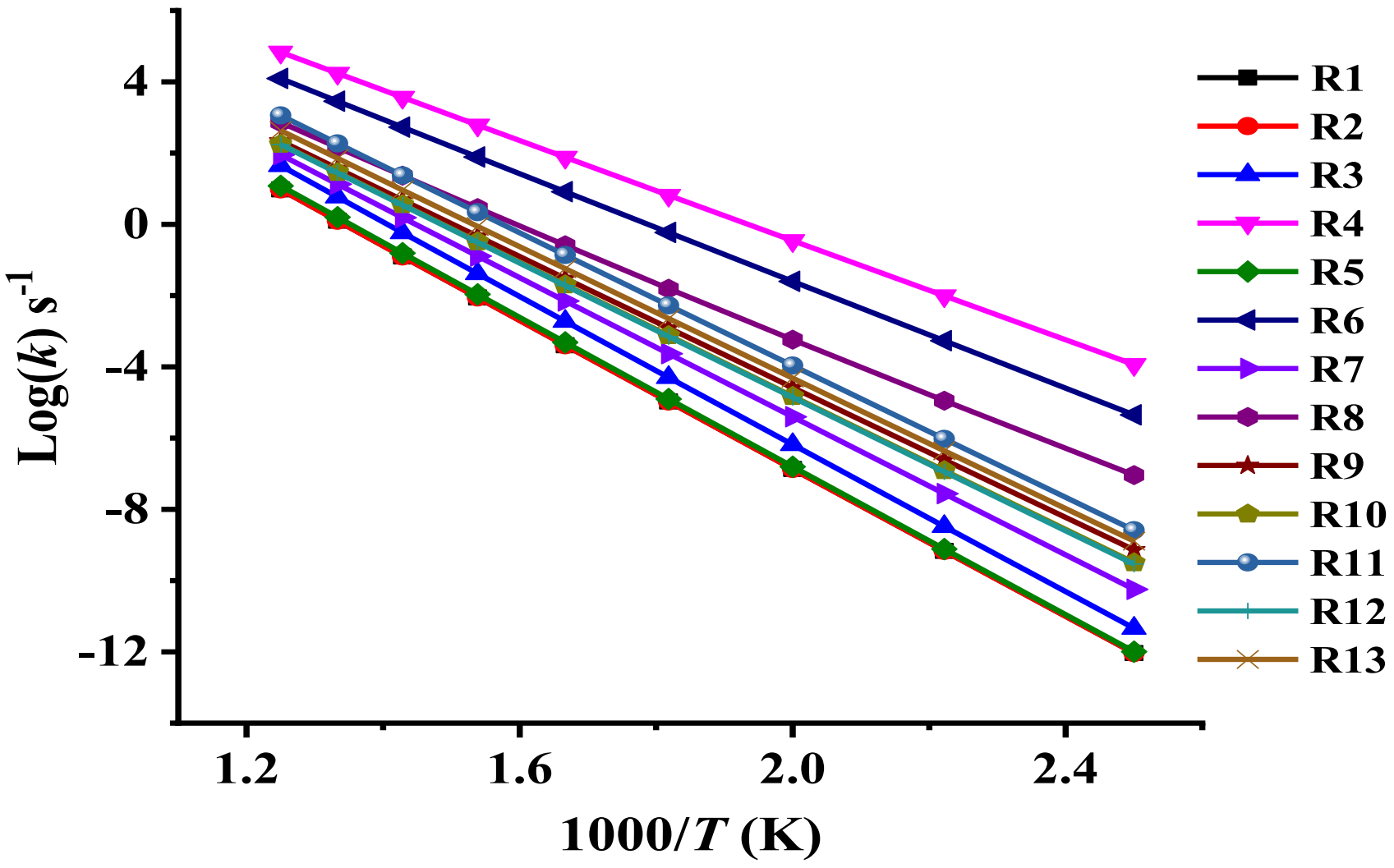

**Fig. 10.** RRKM rate constants for unimolecular decomposition reactions of the investigated ethoxyquinolines and ethoxyisoquinolines (R1- R13) at ($T$ = 400-1200 K, $p$ = 1 bar) of MPW1B95/6-311++G(2d,2p)//BMK/6-31+G(d,p) level of theory.

The pressure dependence of the studied unimolecular H-atom transfer thermal decomposition reactions is calculated employing RRKM theory at a low-pressure range of $10^{-6}$ to 1 bar at 800 K and is sketched in Fig. 6 and summarized in Table S14.

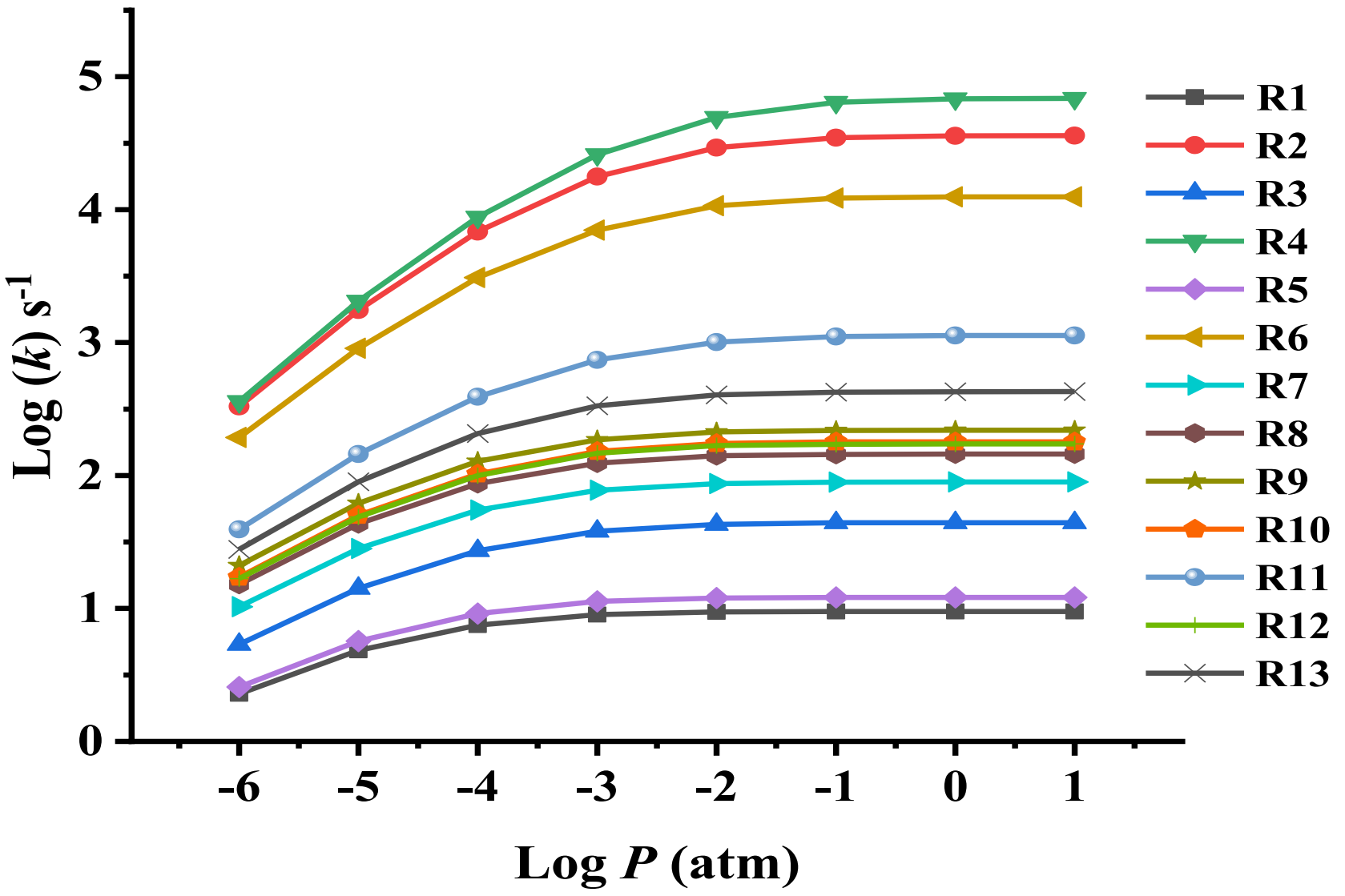

**Fig. 11.** Rate constants $k_{1-13}$ ($s^{-1}$) for unimolecular decomposition reactions of the investigated ethoxyquinolines and ethoxyisoquinolines (R1- R13) obtained from pressure-dependent RRKM theory based on MPW1B95/6-311++G(2d,2p)//BMK/6-31+G(d,p) energies, ($T$ = 800 K, $P$ = 1.0E-06–1.0E+01 bar).

As can be seen in Fig.7, the rate coefficients for reactions (R1-R13) over the applied range of pressure ($10^{-6}$–10 bar), are pressure dependent at the applied temperature.

| **Parameter\ channel** | ***R*₁** | ***R*₂** | ***R*₃** | ***R*₄** | ***R*₅** | ***R*₆** | ***R*₇** |
|---|---|---|---|---|---|---|---|
| Iog$A$($s^{-1}$)<br>aExp | 13.83<br>(13.06))[14] | 13.09 | 14.47<br>(12.92))[14] | 13.47 | 13.99<br>(12.27)[14] | 13.42 | 14.47 |
| $E_a$/kcal/mol<br>aExp | 46.33<br>(47.12))[14] | 30.60 | 46.21<br>(47.66))[14] | 30.96 | 46.52<br>(49.18))[14] | 33.53 | 46.21 |
| $k$($s^{-1}$)<br>aExp | 2.89E+04<br>(2.13E+04))[14] | 4.70E+06 | 1.35E+05<br>(0.20E+05))[14] | 9.89E+06 | 3.86E+04<br>(0.26E+04))[14] | 2.87E+06 | 1.54E+05 |
| **Parameter\ channel** | ***R*₈** | ***R*₉** | ***R*₁₀** | ***R*₁₁** | ***R*₁₂** | ***R*₁₃** | |
| Iog$A$($s^{-1}$)<br>aExp | 13.76 | 13.65 | 13.81 | 14.69 | 13.83 | 14.07<br>(11.28))[16] | |
| $E_a$/kcal/mol | 41.73 | 40.73 | 41.54 | 42.21 | 41.70 | 41.32 | |

| | | | | | | |
|---|---|---|---|---|---|---|
| | | | | | | (30.37))[16] |
| $k(s^{-1})$ [a]Exp | 1.82E+05 | 2.15E+05 | 2.18E+05 | 1.17E+06 | 2.12E+05 | 4.25E+05 (3.41E+05)) [16] |

**Table 3**

Two-parameters Arrhenius coefficients for unimolecular decomposition reactions of the investigated ethoxyquinolines and ethoxyisoquinolines (R1- R13) from TST/Eck calculations ($T$ = 400-1200 K, $p$ =1bar) using MPW1B95/6-311++G(2d,2p) //BMK/6-31+G(d,p) energies. [a]N.A. Al-Awadi et al [14,16].

## 4. CONCLUSIONS

This paper describes the thermochemistry and kinetics of the unimolecular gas-phase thermal decomposition reactions of ten ethoxy- and ethoxyisoquinolines (**1-EisoQ**, **2-EQ**, **3-EQ**, **3-EisoQ**, **4-EQ**, **4-EisoQ**, **5-EQ**, **5-EisoQ**, **8-EQ,** and **8-EisoQ**) at BMK and MPW1B95. The rate constants for these reactions were calculated using conventional transition state theory combined with Eckart tunneling (TST/Eck) and the statistical Rice-Ramsperger-Kassel-Marcus (RRKM) theories. The results obtained can be summarized as follows:

1. Formation of ethylene and keto form is preferred kinetically and thermodynamically.
2. Quinolones and isoquinolones are lower in energy than the corresponding enols except in the case of 3-isoquinolone where there is no aromatic ring exists.
3. The investigated decomposition reactions show clear and significant temperature- and pressure-dependent rate constants in the considered ranges.

**Acknowledgment**

This paper is dedicated to the soul of the late Dr. Ahmed El-Nahas, without whom this project would never have been possible.

**Appendix A. Supplementary material**

Supplementary data associated with this article can be found, in the online version, at